\documentclass[twocolumn,showkeys]{revtex4}
\usepackage[utf8]{inputenc}
\usepackage{graphicx}

\begin{document}

\title{Spin-electron-acoustic waves and solitons in high-density degenerate relativistic plasmas}

\author{Pavel A. Andreev}
\email{andreevpa@physics.msu.ru}
\affiliation{Department of General Physics, Faculty of physics, Lomonosov Moscow State University, Moscow, Russian Federation, 119991.}

\date{\today}

\begin{abstract}
The spin-electron-acoustic waves (sometimes called the spin-plasmons)
can be found in degenerate electron gas
if the spin-up electrons and spin down electrons move relatively each other.
Here, we suggest relativistic hydrodynamics with the separate spin evolution
which allows us to study linear and nonlinear spin-electron-acoustic waves,
including the spin-electron-acoustic solitons.
Presented hydrodynamic model is the corresponding generalization of
the relativistic
hydrodynamic model with the average reverse gamma factor evolution
which
consists of the equations for evolution of the following functions
the partial concentrations (for spin-up electrons and spin down electrons),
the partial velocity fields,
the partial average reverse relativistic gamma factors,
and the partial flux of the reverse relativistic gamma factors.
We find that the relativistic effects decreases the phase velocity of spin-electron-acoustic waves.
Numerical analysis of the changes of spectra of Langmuir wave, spin-electron-acoustic wave, and ion-acoustic wave
under the change of the spin polarization of electrons is presented.
It is demonstrated that spectra of Langmuir wave and spin-electron-acoustic wave getting closer to each other in the relativistic limit.
Spin dependence of the amplitude and width of the relativistic spin-electron-acoustic soliton is demonstrated as well.
Reformation of the bright soliton of potential of the electric field into the dark soliton under the influence of the relativistic effects is found.
\end{abstract}


\maketitle





\section{Introduction}

The spin-electron-acoustic waves (SEAWs) in degenerate plasmas are suggested in 2015-2016
for the three-dimensional, two-dimensional, near surface geometries
of electron-ion, electron-positron, and electron-positron-ion plasmas
\cite{Andreev PRE 15}, \cite{Andreev AoP 15 SEAW}, \cite{Andreev PRE 16}, \cite{Andreev APL 16}, \cite{Andreev EPL 16}.
It it found together with the original model called the separate spin evolution quantum hydrodynamic (SSE-QHD) \cite{Andreev PRE 15},
which is found from the Pauli equation being combination of two equations.
Therefore, the spin-up electrons and spin-down electrons are considered as two different fluids (species or subsystems).
A number of spincaused effects are demonstrated in the SSE-QHD equations including the Euler and continuity equations
which are discussed in Refs.
\cite{Andreev_Iqbal PoP 16},
\cite{Iqbal Shahid PP 17},
\cite{Iqbal PLA 18},
\cite{Trukhanova PLA 15},
\cite{Iqbal PP 18 UHW},
\cite{Iqbal PP 18 LI}.
The separate spin evolution kinetic is developed and used as well
\cite{Andreev PP 16 SSE kin},
\cite{Andreev PoP non-triv kinetics 17},
\cite{Andreev PoP kinetics 17 a},
\cite{Andreev PoP kinetics 17 b}.
Study of the separate spin evolution is part of the field of quantum plasmas developed in the large number of works including
\cite{Maksimov QHM 99},
\cite{MaksimovTMP 2001},
\cite{Marklund PRL07},
\cite{Brodin NJP 07},
\cite{Brodin PRL 08 Cl Reg},
\cite{Brodin PRL 08 g Kin},
\cite{Brodin PRL 10},
\cite{Shukla PhUsp 2010},
\cite{Shukla RMP 11},
\cite{Mahajan PRL 11},
\cite{Koide PRC 13},
\cite{Andreev PP 15 Positrons},
\cite{Yoshida JPA 16},
\cite{Uzdensky RPP review 14},
\cite{Dodin PRA 15 First-principle},
\cite{Ekman PRE 17},
where
the complete quantum hydrodynamic model derivation based on the trace of the microscopic motion of quantum particles
\cite{Maksimov QHM 99},
\cite{MaksimovTMP 2001},
\cite{Andreev Ch 21},
\cite{Andreev PoF 21},
\cite{Andreev JPP 21}.
However, the SEAWs are related to relatively simple quasi-classic manifestation of the spin effects in the Euler equation.
It is the difference of the partial pressures entering the Euler equations spin-up and spin-down electrons.
It opens up a possibility for the partial generalization of the SSE-QHD,
i.e. the generalization of the quasi-classic limit of the SSE-QHD.
This kind of partial generalization is made in this paper.
Here, we present the quasi-classic relativistic SSE-QHD.
This model based on the transition of basic features quasi-classic SSE-QHD up on to classic relativistic
hydrodynamic model with the average reverse gamma factor evolution,
which is recently developed for the degenerate electron gas \cite{Andreev 2021 11}.
The degenerate relativistic
hydrodynamic model with the average reverse gamma factor evolution \cite{Andreev 2021 11} is the result of generalization of
classic relativistic
hydrodynamic model with the average reverse gamma factor evolution
of the relativistically hot plasmas \cite{Andreev 2021 05}, \cite{Andreev 2021 09}, \cite{Andreev 2021 10}.
This model is applied to analysis of spectra of Langmuir, and electromagnetic waves in magnetized relativistic plasmas
including the low frequency dynamics governed by ions
\cite{Andreev 2021 05}, \cite{Andreev 2021 06}, \cite{Andreev 2021 07}, \cite{Andreev 2021 08}.
The derivation of the relativistic hydrodynamic model is based on the original method developed in Refs.
\cite{Kuz'menkov 91}, \cite{Drofa TMP 96}, \cite{Andreev PIERS 2012},
where no probabilistic or statistical assumption is made.
While the derivation is completely based on deterministic classical mechanics.

In order to describe the SSE in the nonrelativistic regime
we present the equations for SSE-QHD following Ref. \cite{Andreev PRE 15}
\begin{equation}\label{SUSD Obl cont eq electrons spin UPandDOWN}
\partial_{t}n_{\uparrow/\downarrow}+\nabla(n_{\uparrow/\downarrow}\textbf{v}_{\uparrow/\downarrow})
=\pm\frac{\mu}{\hbar}(S_{x}B_{y} -S_{y}B_{x}), \end{equation}
where $n_{\uparrow}$ ($n_{\downarrow}$) is the concentration of electrons located in the spin-up (spin-down) state,
$\textbf{v}_{\uparrow}$ ($\textbf{v}_{\downarrow}$) is the velocity field of electrons located spin-up (spin-down).
The right-hand side of the continuity equation contains functions $S_{x}$ and $S_{y}$,
which are projections of the spin density vector,
$\mu$ is the magnetic moment.
The concentration of all electrons $n_{e}=n_{\uparrow}+n_{\downarrow}$ is the sum of partial concentrations.

The transverse projections of spin density evolve in accordance with the following pair of equations
$$\partial_{t}S_{x/y}+\frac{1}{2}\nabla[S_{x/y}(\textbf{v}_{\uparrow}+\textbf{v}_{\downarrow})]$$
\begin{equation}\label{SUSD Obl eq for Sx}
-\frac{\hbar}{4m}\nabla\Biggl([\textbf{S}\times\textbf{e}_{z}]_{x/y}\biggl(\frac{\nabla n_{\uparrow}}{n_{\uparrow}}
-\frac{\nabla n_{\downarrow}}{n_{\downarrow}}\biggr)\Biggr)=T_{x/y},\end{equation}
where
we have no spin-up and spin-down indexes for the spin density functions,
and the following expressions for the spin torque are applied
$T_{x}=\frac{2\mu}{\hbar}(B_{z}S_{y}-B_{y}(n_{u}-n_{d}))$,
and
$T_{y}=\frac{2\mu}{\hbar}(B_{x}(n_{u}-n_{d})-B_{z}S_{x})$.

Finally, we include the pair of vector equations for the partial velocity fields
$$mn_{\uparrow/\downarrow}(\partial_{t}+\textbf{v}_{\uparrow/\downarrow}\nabla)\textbf{v}_{\uparrow/\downarrow}
-\frac{\hbar^{2}}{4m}n_{\uparrow/\downarrow}
\nabla\Biggl(\frac{\triangle n_{\uparrow/\downarrow}}{n_{\uparrow/\downarrow}}
-\frac{(\nabla n_{\uparrow/\downarrow})^{2}}{2n_{\uparrow/\downarrow}^{2}}\Biggr)$$
$$+\nabla p_{\uparrow/\downarrow}
=q_{e}n_{\uparrow/\downarrow}\biggl(\textbf{E}+\frac{1}{c}[\textbf{v}_{\uparrow/\downarrow},\textbf{B}]\biggr)$$
$$+m\frac{\mu}{\hbar}(\textbf{J}_{(M)x}B_{y}-\textbf{J}_{(M)y}B_{x})
-m\textbf{v}_{\uparrow/\downarrow}\frac{\mu}{\hbar}(S_{x}B_{y}-S_{y}B_{x})$$
\begin{equation}\label{SUSD Obl Euler eq electrons spin UPandDOWN}
+\frac{\mu}{2}(S_{x}\nabla B_{x}+S_{y}\nabla B_{y})
\pm\mu n_{\uparrow/\downarrow}\nabla B_{z},\end{equation}
where
\begin{equation}\label{SUSD Obl Spin current XandY}
\textbf{J}_{(M)x/y}=\frac{1}{2}(\textbf{v}_{\uparrow}+\textbf{v}_{\downarrow})S_{x/y}
-\frac{\hbar}{4m} \biggl(\frac{\nabla n_{\uparrow}}{n_{\uparrow}}-\frac{\nabla n_{\downarrow}}{n_{\downarrow}}\biggr)
[\textbf{S}_{y}\times\textbf{e}_{z}]_{x/y} \end{equation}
is the spin current.

The right-hand side of the continuity equations (\ref{SUSD Obl cont eq electrons spin UPandDOWN}) is related to the spin flip.
The second and third group of terms on the right-hand side of the Euler equation
(they are simultaneously proportional to the magnetic moment $\mu$
and magnetic field $\textbf{B}$ with no operators acting on the magnetic moment)
are also related to the spin flip.
This effect is not involved in the quasi-classic relativistic model presented below.
Same is true for the force field acting from nonuniform magnetic field on the spins.
Moreover, the spin precession is neglected below as well.

It would be better to create background of our model being based on the Dirac equation.
However, the quantum hydrodynamic model derived for single electron from the Dirac equation is rather complex
(see Ref. \cite{Asenjo PP 11} for the modern description of this model).
It covers a number of physical effects.
The Dirac equation is developed for the single-electron dynamics
$\gamma^{\mu}(\imath\hbar\partial_{\mu}-qA_{\mu}/c)\psi-mc\psi=0$
which describes the evolution of four component wave bi-spinor function $\psi$,
where we also have
$\gamma^{\mu}$ are the four by four Dirac matrices,
$A^{\mu}$ is the electromagnetic four-potential, and $\mu=0,1,2,3$.

The spin-electron-acoustic waves are theoretically demonstrated in degenerate plasmas
after derivation of the SSE-QHD equations \cite{Andreev PRE 15}
and the derivation of kinetic equations with separate spin evolution \cite{Andreev PP 16 SSE kin},
however the waves of the same nature called the spin-plasmons are suggested
in the condensed matter physics for the two-dimensional electron gas and electron in graphene
\cite{Ryan PRB 91}, \cite{Agarwal PRL 11}, \cite{Agarwal PRB 14}, \cite{Perez PRB 09}.

It is also essential to mention that developed here model of relativistic degenerate electron gas
is novel look on the structure of the relativistic hydrodynamic equations
which are earlier developed for classical
\cite{Hakim book Rel Stat Phys},
\cite{Shatashvili ASS 97}
\cite{Mahajan PRL 03},
\cite{Mahajan PoP 2011},
\cite{Heyvaerts AA 12},
\cite{Asenjo 19},
\cite{Shatashvili PoP 20},
\cite{Comisso PRL 14},
\cite{Liu PPCF 21},
\cite{Comisso 19},
\cite{Mahajan PoP 16},
\cite{Brunetti MNRAS 04},
and quantum plasmas
\cite{Uzdensky RPP review 14},
\cite{Asenjo PP 11},
\cite{Mahajan IJTP 14},
\cite{Mendonca PP 11},
\cite{Zhu PPCF 12},
\cite{Munoz EPS 06}.

This paper is organized as follows.
In Sec. II the relativistic hydrodynamic model with the average reverse gamma factor evolution
is adopted for the separate spin evolution in systems of degenerate partially spin polarized electrons.
In Sec. III single fluid reduction of the relativistic separate spin evolution hydrodynamics.
In Sec. IV the linear and nonlinear analysis of the small amplitude spin-electron-acoustic waves is given.
In Sec. V a brief summary of obtained results is presented.

\section{Two fluid relativistic hydrodynamic model with the average reverse gamma factor evolution}

In this paper we present the separate spin evolution relativistic hydrodynamics
as the two fluid hydrodynamic model with the average reverse gamma factor evolution.
This model is based on generalization of models developed in Refs. \cite{Andreev PRE 15}, \cite{Andreev 2021 11},
\cite{Andreev 2021 05}, \cite{Andreev 2021 09}.

We start presentation of the suggested separate spin evolution model with the continuity equation
\begin{equation}\label{RHD2021ClLM cont via v} \partial_{t}n_{s}+\nabla\cdot(n_{s}\textbf{v}_{s})=0,\end{equation}
which has rather traditional form,
it is also corresponds to Refs. \cite{Andreev 2021 11}, \cite{Andreev 2021 05}, \cite{Andreev 2021 09}.
However, we consider the separate spin evolution,
so we should present comparison with the separate spin evolution quantum hydrodynamics obtained from the Pauli equation
\cite{Andreev PRE 15}, \cite{Andreev AoP 15 SEAW}.
It shows that the continuity equations for the partial concentrations $n_{\uparrow}$ and $n_{\downarrow}$
have non zero right-hand side.
The spin polarization of electrons changes under action of the magnetic field.
It corresponds to the change of difference of partial concentrations $n_{\uparrow}$ and $n_{\downarrow}$ and corresponding change of each of them
caused by the torque created by magnetic field and acting on the magnetic moments of electrons.
More exactly, we have the z-projection of the torque
while the z-direction is the chosen direction and we consider the spin-projections relatively this direction.
The z-projection of the torque appears to be nonlinear,
so we have no restrictions at the analysis of the linear effects.
The nonlinear analysis of the longitudinal perturbations has also zero contribution of the torque.
Therefore, the model under consideration allows to consider a wide range of phenomena.

The second equation of evolution is the Euler equation is given
which is the velocity field evolution equation \cite{Andreev 2021 11},
\cite{Andreev 2021 05}, \cite{Andreev 2021 09}:
$$n_{s}\partial_{t}\textbf{v}_{s}+n_{s}(\textbf{v}_{s}\cdot\nabla)\textbf{v}_{s}
+\frac{1}{m_{s}}\nabla\tilde{p}_{s}$$
$$=\frac{q_{s}}{m_{s}}\biggl(\Gamma_{s} -\frac{\tilde{t}_{s}}{c^{2}}\biggr)\textbf{E}
+\frac{q_{s}}{m_{s}c}[(\Gamma_{s} \textbf{v}_{s}+\textbf{t}_{s})\times\textbf{B}]$$
\begin{equation}\label{RHD2021ClLM Euler for v}
-\frac{q_{s}}{m_{s}c^{2}}
\biggl(\Gamma_{s} \textbf{v}_{s} (\textbf{v}_{s}\cdot \textbf{E})
+\textbf{v}_{s} (\textbf{t}_{s}\cdot\textbf{E})+\textbf{t}_{s} (\textbf{v}_{s}\cdot\textbf{E})\biggr), \end{equation}
where
$m_{s}$ and $q_{s}$ are the mass and charge of particle of $s$ species,
$c$ is the speed of light,
tensor $p^{ab}_{s}=\tilde{p}_{s}\delta^{ab}$
is the flux of the velocities for electrons with fixed spin projection,
tensor $t^{ab}_{s}=\tilde{t}_{s}\delta^{ab}$
is the flux of the average reverse gamma-factor for spin-s electrons,
$\delta^{ab}$ is the three-dimensional Kronecker symbol.
Moreover, we work in the Minkovskii space, hence the metric tensor has diagonal form
with the following sings $g^{\alpha\beta}=\{-1, +1, +1, +1\}$.
We mostly use the three dimensional notations,
therefore, we can change covariant and contrvariant indexes for the three-vector indexes: $v^{b}_{s}=v_{b,s}$.
The Latin indexes like $a$, $b$, $c$ etc describe the three-vectors,
while the Greek indexes are deposited for the four-vector notations.
We also have Latin index $s$ which refers to the species or subspecies of electrons with different spin projections.
However, the indexes related to coordinates are chosen from the beginning of the alphabet,
while other indexes are chosen in accordance with their physical meaning.
The model under presentation and this Euler equation includes no effects related to change of the spin projections of electrons on the chosen direction.

The equation of evolution of the averaged reverse relativistic gamma factor for electrons with fixed spin projection in absence of the spin-flip effects has the following form
$$\partial_{t}\Gamma_{s}+\nabla(\Gamma_{s} \textbf{v}_{s}+\textbf{t}_{s})$$
\begin{equation}\label{RHD2021ClLM eq for Gamma}
=-\frac{q_{s}}{m_{s}c^{2}}n(\textbf{v}_{s}\cdot\textbf{E})
\biggl(1-\frac{1}{c^{2}}\biggl(\textbf{v}_{s}^{2}+\frac{5\tilde{p}_{s}}{n_{s}}\biggr)\biggr).\end{equation}
Function $\Gamma_{s}$ is the hydrodynamic Gamma function \cite{Andreev 2021 05} considered here for electrons with particular spin projection.


The model also includes the equation of evolution for the current of the reverse relativistic gamma factor
(the hydrodynamic Theta function) presented for electrons with fixed spin projection:
$$(\partial_{t}+\textbf{v}_{s}\cdot\nabla)\textbf{t}_{s}^{a}
+\nabla ^{a}\tilde{t}_{s}
+(\textbf{t}_{s}\cdot\nabla) \textbf{v}_{s}^{a}+\textbf{t}_{s}^{a} (\nabla\cdot \textbf{v}_{s})$$
$$+\Gamma_{s}(\partial_{t}+\textbf{v}_{s}\cdot\nabla)\textbf{v}_{s}^{a}
=\frac{q_{s}}{m_{s}}n_{s}\textbf{E}^{a}\biggl[1-\frac{\textbf{v}_{s}^{2}}{c^{2}}-\frac{3\tilde{p}_{s}}{n_{s}c^{2}}\biggr]$$
$$+\frac{q_{s}}{m_{s}c}[n_{s}\textbf{v}_{s}\times \textbf{B}]
\biggl[1-\frac{\textbf{v}_{s}^{2}}{c^{2}}-\frac{5\tilde{p}_{s}}{n_{s}c^{2}}\biggr]
-\frac{2q_{s}}{m_{s}c^{2}}\Biggl[\textbf{E}^{a}\tilde{p}_{s}\biggl(1-\frac{\textbf{v}_{s}^{2}}{c^2}\biggr)$$
\begin{equation}\label{RHD2021ClLM eq for t a}
+n_{s}\textbf{v}_{s}^{a}(\textbf{v}_{s}\cdot\textbf{E})\biggl(1-\frac{\textbf{v}_{s}^{2}}{c^{2}}-\frac{9\tilde{p}_{s}}{n_{s}c^{2}}\biggr)
-\frac{5M_{s0}}{3c^{2}} \textbf{E}^{a}\Biggr].\end{equation}
The hydrodynamic equations
(\ref{RHD2021ClLM cont via v}), (\ref{RHD2021ClLM Euler for v}), (\ref{RHD2021ClLM eq for Gamma}) and (\ref{RHD2021ClLM eq for t a})
are obtained in the mean-field approximation (the self-consistent field approximation).
The fourth rank tensor $M^{abcd}_{s}$ existing in equation (\ref{RHD2021ClLM eq for t a})
is presented via its partial trace $M^{abcc}_{s}=M_{s,c}^{cab}$.
This tensor is constructed via the Kronecker symbols
$M^{abcd}_{s}=(M_{s0}/3)(\delta^{ab}\delta^{cd}+\delta^{ac}\delta^{bd}+\delta^{ad}\delta^{bc})$.
So we have the following elements of this tensor: $M_{s}^{xxxx}=M_{s}^{yyyy}=M_{s}^{zzzz}=M_{s0}$
and $M_{s}^{xxyy}=M_{s}^{xxzz}=M_{s0}/3$.
Otherwise the element of tensor $M^{abcd}_{s}$ is equal to zero.
Required partial trace $M_{s,c}^{cab}$ has the following form $M_{s,c}^{cab}=(5M_{s0}/3)\delta^{ab}$.

The equations of electromagnetic field have the traditional form
presented in the three-dimensional notations $\nabla \cdot\textbf{B}=0$,
\begin{equation}\label{RHD2021ClLM rot E} \nabla\times \textbf{E}=-\frac{1}{c}\partial_{t}\textbf{B},\end{equation}
\begin{equation}\label{RHD2021ClLM div E with time} \nabla \cdot\textbf{E}=
4\pi(q_{e}n_{e\uparrow}+q_{e\downarrow}n_{e}+q_{i}n_{i}),
\end{equation}
and
\begin{equation}\label{RHD2021ClLM rot B with time}
\nabla\times \textbf{B}=\frac{1}{c}\partial_{t}\textbf{E}+
\frac{4\pi q_{e}}{c}n_{e\uparrow}\textbf{v}_{e\uparrow}
+\frac{4\pi q_{e}}{c}n_{e\downarrow}\textbf{v}_{e\downarrow}
+\frac{4\pi q_{i}}{c}n_{i}\textbf{v}_{i},
\end{equation}
where the ions exist as the motionless background.

\subsection{Equations of state for spin-up and spin-down electrons}

Equations (\ref{RHD2021ClLM cont via v})-(\ref{RHD2021ClLM eq for t a}) are derived for the relativistically hot plasmas
\cite{Andreev 2021 05}, \cite{Andreev 2021 09}.
Next, it is adopted for the degenerate electron gas \cite{Andreev 2021 11}.
Our goal here is further generalization of the hydrodynamic model with the average reverse gamma factor evolution
to consider SEAW \cite{Andreev PRE 15}, \cite{Andreev PRE 16},
so we need to include the separate spin evolution to this model.
We consider the degenerate electron gas of high concentration,
hence we find the Fermi velocity $v_{Fs}=p_{Fs}/\sqrt{1+p_{Fs}^{2}/m_{s}^{2}c^{2}}m_{s}$,
where $p_{Fs}=(6\pi^{2}n_{s})^{1/3}\hbar$.

For the analysis of the high density relativistic degenerate electron gas subspecies with the chosen spin projection
we derive the equations of state for functions
$p_{s}^{ab}$, $t_{s}^{ab}$, and $M_{s}^{abcd}$.

Degenerate electrons with the fixed spin projection are described within the zero-temperature limit of the Fermi-Dirac distribution,
which is given by the Fermi step distribution
\begin{equation}\label{RHD2021ClLM Fermi step} f_{s0}=\Biggl\{\begin{array}{c}
                                                               \frac{1}{(2\pi\hbar)^{3}} \\
                                                               0
                                                             \end{array}
\textrm{for}
\begin{array}{c}
                                                               p\leq p_{Fs} \\
                                                               p> p_{Fs}
                                                             \end{array}
\end{equation}

The concentration has well-known form in terms of the distribution function
\begin{equation}\label{RHD2021ClLM concentr via f} n_{s}=\int f_{s0} d^{3}p. \end{equation}
where
$\textbf{p}=m_{s}\textbf{v}/\sqrt{1-\textbf{v}^{2}/c^{2}}$.
The flux of the current of particles,
which has the following representation in form of the distribution function
\begin{equation}\label{RHD2021ClLM p via f} p_{s}^{ab}=\int v^{a}v^{b} f_{s0} d^{3}p. \end{equation}
We substitute distribution function (\ref{RHD2021ClLM Fermi step}) go calculate
the equation of state $p_{s}^{ab}=\tilde{p}_{s} \delta^{ab}$ with
\begin{equation}\label{RHD2021ClLM p rel eq of state}
\tilde{p}_{s}=\frac{m_{s}^{3}c^{5}}{6\pi^{2}\hbar^{3}}\biggl[\frac{1}{3}\xi_{s}^{3}-\xi_{s}+\arctan\xi_{s}\biggr],
\end{equation}
where
$\xi_{s}\equiv p_{Fs}/mc$.
The flux of the current of the average reverse gamma factor can also be presented via the distribution function
\begin{equation}\label{RHD2021ClLM t via f} t_{s}^{ab}=\int \biggl(\frac{v^{a}v^{b}}{\gamma}\biggr) f_{s0} d^{3}p. \end{equation}
We obtain $t^{ab}_{s}=\tilde{t}_{s} \delta^{ab}$,
where
\begin{equation}\label{RHD2021ClLM t rel eq of state}
\tilde{t}_{s}=\frac{m_{s}^{3}c^{5}}{12\pi^{2}\hbar^{3}} \biggl[ \xi_{s}\sqrt{\xi_{s}^{2}+1}+\frac{2\xi_{s}}{\sqrt{\xi_{s}^{2}+1}} -3Arsinh\xi_{s}\biggr],
\end{equation}
with
$Arsinh\xi=ln\mid \xi+\sqrt{\xi^{2}+1}\mid$,
and
$sinh(Arsinh\xi)=\xi$.
The fourth rank tensor $M_{s}^{abcd}$ is also calculated
\begin{equation}\label{RHD2021ClLM M via f} M_{s}^{abcd}=\int v^{a}v^{b} v^{c}v^{d} f_{s0} d^{3}p, \end{equation}
to get the following expression
$M_{s}^{abcd}=(M_{s0}/3)(\delta^{ab}\delta^{cd}+\delta^{ac}\delta^{bd}+\delta^{ad}\delta^{bc})$,
where
\begin{equation}\label{RHD2021ClLM M rel eq of state}
M_{s0}=\frac{m_{s}^{3}c^{7}}{60\pi^{2}\hbar^{3}} \biggl[ 2\xi_{s}(\xi_{s}^{2} -6) -\frac{3\xi_{s}}{\xi_{s}^{2}+1} +15\arctan\xi_{s} \biggr].
\end{equation}
We also need to find the equilibrium expression for the partial average reverse gamma factor $\Gamma_{0s}$ for the spin-s degenerate electrons
\begin{equation}\label{RHD2021ClLM Gamma via f} \Gamma_{s}=\int \frac{1}{\gamma} f_{s0} d^{3}p, \end{equation}
where we find
\begin{equation}\label{RHD2021ClLM Gamma rel eq of state}
\Gamma_{0s}= \frac{m_{s}^{3}c^{3}}{4\pi^{2}\hbar^{3}} \biggl[ \xi_{s}\sqrt{\xi_{s}^{2}+1} -Arsinh\xi_{s}\biggr].
\end{equation}
These expressions leads to the closed set of hydrodynamic equations,
which are applied below to the longitudinal waves,
particularly the spin-electron-acoustic waves
\cite{Andreev PRE 15}, \cite{Andreev PRE 16}, \cite{Andreev APL 16}, \cite{Andreev JPP 21}.

\section{Single fluid reduction of separate spin evolution relativistic hydrodynamic model with the average reverse gamma factor evolution}

Spin polarized system of electrons can be described as the single fluid
\cite{MaksimovTMP 2001}, \cite{Marklund PRL07}, \cite{Mahajan PRL 11},
where the spin density exists among other well-known hydrodynamic functions like the concentration and the velocity field.
Method of transition from the separate spin evolution quantum hydrodynamics to the single fluid quantum hydrodynamics of electrons is described
in Ref. \cite{Andreev AoP 15 SEAW} for the nonrelativistic regime.
Here we apply this method for the relativistic separate spin evolution quasi-classic hydrodynamics described above.

The superposition of partial concentrations $n_{s}$ gives the full concentration of electrons:
$n=n_{\uparrow}+n_{\downarrow}$.
The flux of electrons is the superposition of partial fluxes,
so it gives us relation between velocities:
$\textbf{v}=\frac{n_{\uparrow}\textbf{v}_{\uparrow}+n_{\downarrow}\textbf{v}_{\downarrow}}{n_{\uparrow}+n_{\downarrow}}$.
Superposition of the velocity evolution equations (\ref{RHD2021ClLM Euler for v})
shows the structure of the full flux of particle current $\Pi^{ab}$ constructed as the sum of spin-s fluxes of particle current
$\Pi^{ab}_{s}=n_{s}v_{s}^{a}v_{s}^{b}+p_{s}^{ab}$:
$\Pi^{ab}=\Pi^{ab}_{u}+\Pi^{ab}_{d}$.
It contains the contribution of flux on the difference of partial velocity fields:
$\Pi^{ab}=nv^{a}v^{b}+p^{ab}+
\frac{n_{u}n_{d}}{n}(v_{u}^{a}-v_{d}^{a})(v_{u}^{b}-v_{d}^{b})$.
The last term gives transition of center of mass of electrons at the relative motion of the spin-up and spin-down electrons.
Hence, if the difference of velocity fields for the spin-up and spin-down electrons is large
we cannot make transition to the single fluid dynamics.
However, if this difference is small in compare with the velocity field $v^{a}$
we can consider the single fluid hydrodynamics of the spin polarized degenerate electrons.

The average reverse gamma factor is the sum of the partial functions:
$\Gamma=\Gamma_{\uparrow}+\Gamma_{\downarrow}$.
While the flux of average reverse gamma factor has additional contribution related to different partial velocity fields
$\textbf{t}=\textbf{t}_{\uparrow}+\textbf{t}_{\downarrow}
+(\frac{n_{\downarrow}\Gamma_{\uparrow}-n_{\uparrow}\Gamma_{\downarrow}}{n_{\uparrow}+n_{\downarrow}})(\textbf{v}_{\uparrow}-\textbf{v}_{\downarrow})$.

The effects of nonzero spin polarization in the single fluid can be described by the following parameter
$\eta=\mid n_{\uparrow}-n_{\downarrow}\mid/(n_{\uparrow}+n_{\downarrow})$.
We use to present the equations of state for the high-rank tensors entering
the single fluid relativistic hydrodynamics of the spin-polarized degenerate electrons.

The flux of current of particles
is found for the single fluid regime using expression (\ref{RHD2021ClLM p rel eq of state})
$$\tilde{p}=\tilde{p}_{\uparrow}+\tilde{p}_{\downarrow}=\frac{m^{3}c^{5}}{6\pi^{2}\hbar^{3}}
\biggl[\frac{2}{3}\xi^{3}-\biggl((1+\eta)^{1/3}+(1-\eta)^{1/3}\biggr)\xi$$
\begin{equation}\label{RHD2021ClLM p rel eq of state SF}
+\arctan(1+\eta)^{1/3}\xi+\arctan(1-\eta)^{1/3}\xi\biggr],
\end{equation}
where
the first term on the right hand-side $\xi^{3}$ has no spin dependent coefficient except coefficient 2 appearing from
superposition of $(1+\eta)$ and $(1-\eta)$,
and
$\xi\equiv p_{Fe}/mc$,
with the Fermi momentum $p_{Fe}=(3\pi^{2}n_{0})^{1/3}\hbar$.

The single fluid
flux of the current of the average
reverse gamma factor at the nonzero spin polarization appears as the superposition of partial functions
given by equation (\ref{RHD2021ClLM t rel eq of state})
$$\tilde{t}=\tilde{t}_{\uparrow}+\tilde{t}_{\downarrow}
=\frac{m^{3}c^{5}}{12\pi^{2}\hbar^{3}}\times$$
$$\times\Biggl[
\frac{2(1+\eta)^{1/3}\xi}{\sqrt{(1+\eta)^{2/3}\xi^{2}+1}}
+\frac{2(1-\eta)^{1/3}\xi}{\sqrt{(1-\eta)^{2/3}\xi^{2}+1}}$$
$$+(1+\eta)^{1/3}\xi
\sqrt{(1+\eta)^{2/3}\xi^{2}+1}$$
$$+(1-\eta)^{1/3}\xi
\sqrt{(1-\eta)^{2/3}\xi^{2}+1}$$
\begin{equation}\label{RHD2021ClLM t rel eq of state SF}
-3Arsinh(1+\eta)^{1/3}\xi-3Arsinh(1-\eta)^{1/3}\xi\Biggr].
\end{equation}
Equation (\ref{RHD2021ClLM t rel eq of state SF}) is the generalization of the expression found in Ref. \cite{Andreev 2021 11}
up to the account of nonzero spin polarization.

Same generalization is made for the function $M_{0}$:
$$M_{0}=M_{\uparrow0}+M_{\downarrow0}
=\frac{m^{3}c^{7}}{60\pi^{2}\hbar^{3}}\times$$
$$\times\Biggl[ 15\arctan(1+\eta)^{1/3}\xi+15\arctan(1-\eta)^{1/3}\xi$$
$$+2(1+\eta)^{1/3}\xi
[(1+\eta)^{2/3}\xi^{2} -6]$$
$$+2(1-\eta)^{1/3}\xi
[(1-\eta)^{2/3}\xi^{2} -6]$$
\begin{equation}\label{RHD2021ClLM M rel eq of state SF}
-\frac{3(1+\eta)^{1/3}\xi}{(1+\eta)^{2/3}\xi^{2}+1}
-\frac{3(1-\eta)^{1/3}\xi}{(1-\eta)^{2/3}\xi^{2}+1}
\Biggr].
\end{equation}

Finally, we find expression for the equilibrium average reverse gamma factor at the nonzero spin polarization
$$\Gamma_{0}=\Gamma_{0\uparrow}+\Gamma_{0\downarrow}= \frac{m^{3}c^{3}}{4\pi^{2}\hbar^{3}}\times$$
$$\times\biggl[ (1+\eta)^{1/3}\xi
\sqrt{(1+\eta)^{2/3}\xi^{2}+1}$$
$$+(1-\eta)^{1/3}\xi
\sqrt{(1-\eta)^{2/3}\xi^{2}+1}$$
\begin{equation}\label{RHD2021ClLM Gamma rel eq of state SF}
-Arsinh(1+\eta)^{1/3}\xi+Arsinh(1-\eta)^{1/3}\xi\biggr].
\end{equation}

Results presented in this section allows to partially consider the effects of the spin polarization in terms of single fluid hydrodynamic model.
However, the spin-electron-acoustic discussed below can not appear in the single fluid model of electrons.

\section{Spin-electron-acoustic waves in the relativistic magnetized partially spin polarized plasmas}

\subsection{Linearized equations for relativistic separate spin evolution hydrodynamics}

We focus on degenerate electron-ion plasmas,
where both components are degenerate.
Moreover, the concentrations of both components $n_{0e}=n_{0i}$ are equal.
Nevertheless, the concentration is large enough,
so the Fermi velocity $v_{Fe}$ getting close to the speed of light $c$.

All species are assumed to be macroscopically motionless in the equilibrium state
$\textbf{v}_{0\uparrow}=\textbf{v}_{0\downarrow}=\textbf{v}_{0i}=0$.
The macroscopic equilibrium electric field is equal to zero either.
Wave propagate parallel to the constant uniform external magnetic field,
so ne effects except the spin polarization are caused by the magnetic field.
The spin polarization manifests itself in the equilibrium in the difference of the partial concentrations of electrons
$n_{0\uparrow}\neq n_{0\downarrow}$.

We consider small amplitude plane wave perturbations of the equilibrium state.
Two linearized continuity equation are found for spin-up and spin-down electrons
\begin{equation}\label{RHD2021ClLM continuity equation lin 1D}
\partial_{t}\delta n_{s}+n_{0s}\partial_{x} \delta v_{xs}=0, \end{equation}
where $s=\uparrow$, $\downarrow$, while ions are assumed to be motionless, so we do not present equations of motion for ions.
Below we include the motion of ions.
To include the motion of ions we assume that parameter $s$ has three values
$\uparrow$ for the spin-up electrons,
$\downarrow$ for the spin-down electrons,
and
$i$ for ions.
So, we do not need to repeat the set of linearized hydrodynamic equations for the second regimes of mobile ions.

The linearized equations for the evolution of the partial velocity fields appear from equation (\ref{RHD2021ClLM Euler for v})
\begin{equation}\label{RHD2021ClLM velocity field evolution equation lin 1D}
n_{0s}\partial_{t}\delta v_{xs}+\frac{\delta \tilde{p}_{0s}}{\delta n_{0s}}\partial_{x}n_{s}
=\frac{q_{s}}{m_{s}}\Gamma_{0s} \delta E_{x}-\frac{q_{s}}{m_{s}c^{2}}\tilde{t}_{0s}\delta E_{x},
\end{equation}
where parameters $\Gamma_{0s}$, $\delta \tilde{p}_{0s}/\delta n_{0s}$ and $\tilde{t}_{0s}$ appear from equations of state presented above
(\ref{RHD2021ClLM Gamma rel eq of state}), (\ref{RHD2021ClLM p rel eq of state}), and (\ref{RHD2021ClLM t rel eq of state}), correspondingly.

The third and fourth pairs of equations in the set of relativistic linearized equations are obtained for evolution of $\delta\Gamma$ and $\delta t_{x}$.
They appear from equations (\ref{RHD2021ClLM eq for Gamma})
and (\ref{RHD2021ClLM eq for t a}).
They have the following form
\begin{equation}\label{RHD2021ClLM evolution of Gamma lin 1D}
\partial_{t}\delta\Gamma_{s} +\Gamma_{0s}\partial_{x}\delta v_{xs}+\partial_{x}\delta t_{xs} =0, \end{equation}
and
$$\partial_{t}\delta t_{xs} +\partial_{x}\delta \tilde{t}_{s}-\frac{\Gamma_{0s}}{n_{0s}}\partial_{x}\delta \tilde{p}_{s}
+\frac{q_{s}}{m_{s}}\frac{\Gamma_{0s}^{2}}{n_{0s}}\delta E_{x}$$
\begin{equation}\label{RHD2021ClLM evolution of Theta lin 1D}
=\frac{q_{s}}{m_{s}}n_{0s}\delta E_{x} -\frac{5q_{s}}{m_{s}c^{2}}\tilde{p}_{0s}\delta E_{x} +\frac{10q_{s}}{3m_{s}c^{2}}M_{0s}\delta E_{x}, \end{equation}
where $M_{0s}^{xxcc}=(5/3)M_{0s}$.
However, equations (\ref{RHD2021ClLM evolution of Gamma lin 1D}) and (\ref{RHD2021ClLM evolution of Theta lin 1D})
do not give any contribution in the spectra of longitudinal waves propagating parallel to the external magnetic field.

We need to consider one of the Maxwell equations for divergence of electric field
since we study the longitudinal waves propagating parallel to the external magnetic field.
The linearized Poisson equation has the well-known form
\begin{equation}\label{RHD2021ClLM Poisson equation lin}
\partial_{x}\delta E_{x}=4\pi (q_{e} \delta n_{e}+q_{i} \delta n_{i}). \end{equation}

\begin{figure} \includegraphics[width=8cm,angle=0]{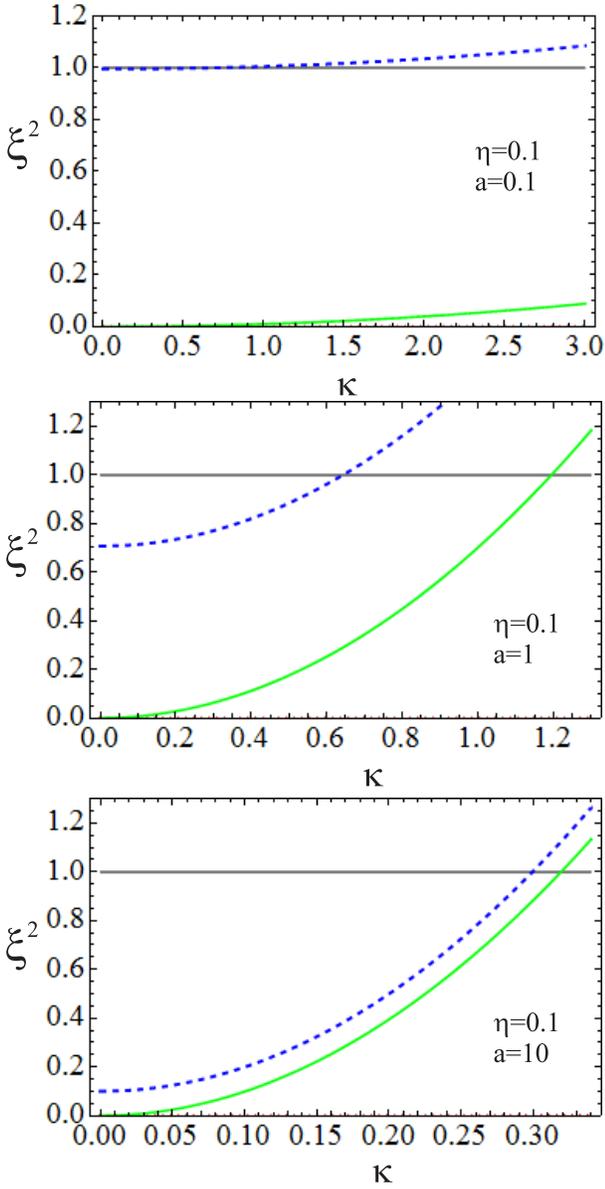}
\caption{\label{RHD2021ClLM Fig 01}
The spectra for the Langmuir wave (dashed blue line), and SEAW (continuous green line) are demonstrated
via plotting of the dimensionless square of frequency $\xi=\omega/\omega_{Le}$
as the function of the dimensionless wave vector $\kappa=kc/\sqrt{3}\omega_{Le}$.
Figures are made for chosen spin polarization $\eta=0.1$,
but for different concentrations of electrons.
The concentrations are presented via dimensionless parameter $a=(3\pi^{2}n_{0e})^{1/3}\hbar/m_{e}c$.
All figures of spectra contain the ion-acoustic wave spectrum.
However, it is almost invisible red dotted line near zero on this scale.
Horisontal line $\xi=1$ corresponds to $\omega=\omega_{Le}$.
}\end{figure}

\begin{figure} \includegraphics[width=8cm,angle=0]{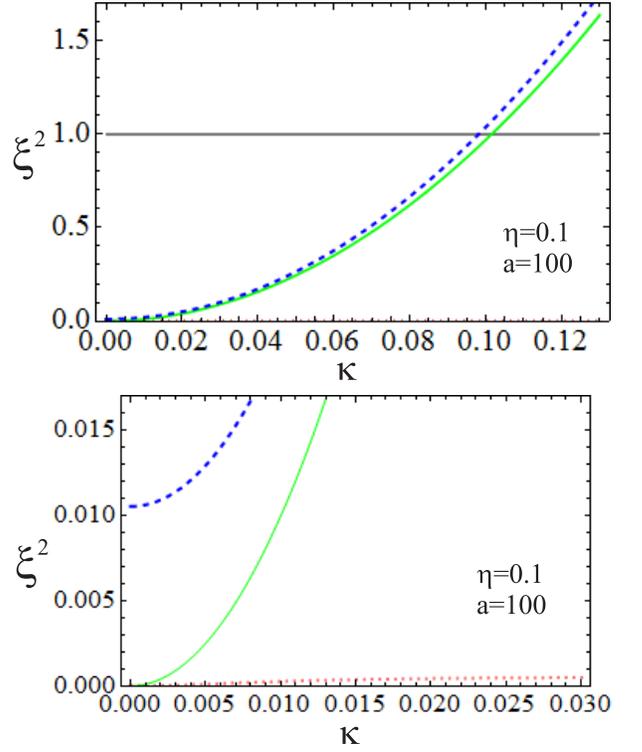}
\caption{\label{RHD2021ClLM Fig 02}
The spectra for the Langmuir wave, SEAW, and ion-acoustic wave (red dotted line) are demonstrated
in the ultrarelativistic limit $a=100$ at $\eta=0.1$.
The upper figure shows the spectra on the large scale.
The lower figure shows the small frequency part of the spectra.
}\end{figure}

\begin{figure} \includegraphics[width=8cm,angle=0]{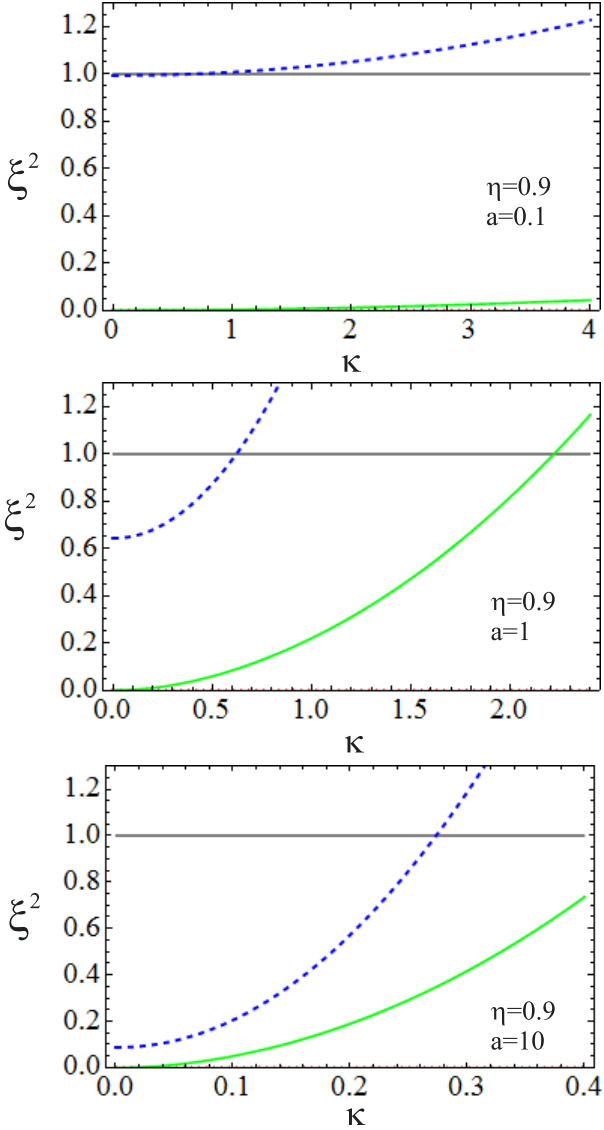}
\caption{\label{RHD2021ClLM Fig 03}
The spectra for the Langmuir wave, and SEAW are presented for chosen spin polarization $\eta=0.9$.
}\end{figure}

\begin{figure} \includegraphics[width=8cm,angle=0]{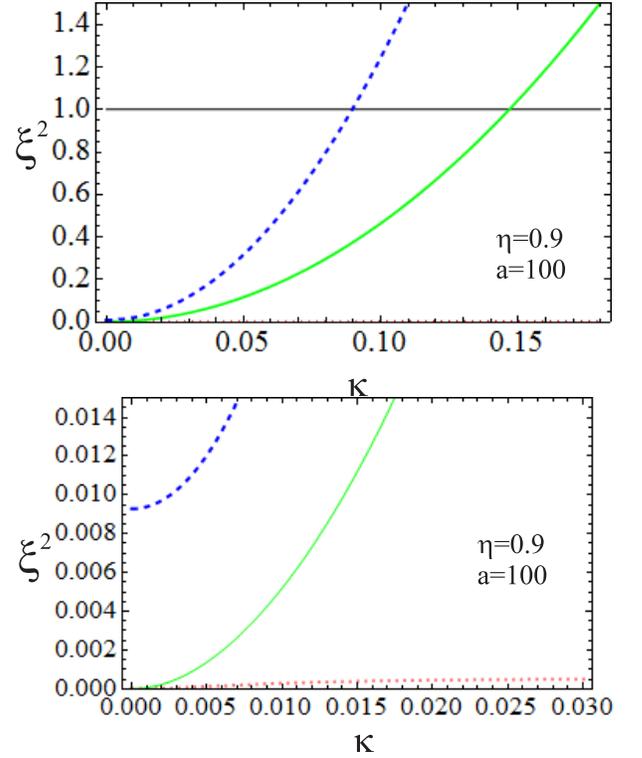}
\caption{\label{RHD2021ClLM Fig 04}
The spectra for the Langmuir wave, SEAW, and ion-acoustic wave are shown in the ultrarelativistic limit at $\eta=0.9$.
}\end{figure}


\subsection{Spectra of the Langmuir waves, spin-electron-acoustic waves, and ion-acoustic waves}

If we consider the relativistic Langmuir waves in the zero-spin-polarization single fluid model of degenerate electrons
we find \cite{Andreev 2021 11}
\begin{equation}\label{RHD2021ClLM Langmuir wave H}
\omega^{2}=\frac{\omega_{Le}^{2}}{\gamma_{Fe}} +\frac{1}{3}v_{Fe}^{2}k_{z}^{2}, \end{equation}
where $\frac{\Gamma_{0}}{n_{0}}-\frac{u_{t}^{2}}{c^{2}}=\frac{1}{\gamma_{Fe}}$,
and
$v_{Fe}^{2}=c^{2}\frac{p_{Fe}^{2}}{p_{Fe}^{2}+m^{2}c^{2}}=\frac{p_{Fe}^{2}}{m^{2}\gamma_{Fe}^{2}}$,
with
$p_{Fe}=(3\pi^{2}n_{0e})^{1/3}\hbar$,
and
$\gamma_{Fe}=1/\sqrt{1-v_{Fe}^{2}/c^{2}}=\sqrt{1+p_{Fe}^{2}/m^{2}c^{2}}$
is the standard relativistic gamma factor considered for the Fermi velocity.

Equations
(\ref{RHD2021ClLM continuity equation lin 1D})-(\ref{RHD2021ClLM Poisson equation lin})
allows us to find the dispersion equation for the high-frequency longitudinal waves
\begin{equation}\label{RHD2021ClLM Disp eq u-d}
1=\frac{1}{\gamma_{F\uparrow}}\frac{\omega_{L\uparrow}^{2}}{\omega^{2}-u_{p\uparrow}^{2}k_{z}^{2}}
+\frac{1}{\gamma_{F\downarrow}}\frac{\omega_{L\downarrow}^{2}}{\omega^{2}-u_{p\downarrow}^{2}k_{z}^{2}}, \end{equation}
where the characteristic velocities $u_{ps}$ have simple expressions via the partial Fermi velocities $u_{ps}^{2}=v_{Fs}^{2}/3$.

For the intermediate frequency regime $v_{F\downarrow}^{2}k_{z}^{2}/3\gg\omega^{2}\gg v_{F\uparrow}^{2}k_{z}^{2}/3$
considered in the long-wavelength limit $k_{z}\rightarrow0$
equation (\ref{RHD2021ClLM Disp eq u-d}) simplifies to the spectrum of the SEAWs
\begin{equation}\label{RHD2021ClLM spectrum seaw small k}
\omega^{2}
=\frac{n_{0\uparrow}}{n_{0\downarrow}}
\frac{\gamma_{F\downarrow}}{\gamma_{F\uparrow}}v_{F\downarrow}^{2}k_{z}^{2}/3,
\end{equation}
or
this expression can be represented in order to explicitly show the dependencies on the spin polarization and concentration of all electrons
\begin{equation}\label{RHD2021ClLM spectrum seaw small k 2}
\omega^{2}=\frac{1-\eta}{(1+\eta)^{1/3}} \frac{(3\pi^{2})^{2/3}\hbar^{2}n_{0e}^{2/3}k_{z}^{2}}{3m_{e}^{2}\sqrt{1+(1+\eta)^{2/3}\frac{(3\pi^{2})^{2/3}\hbar^{2}n_{0e}^{2/3}}{m_{e}^{2}c^{2}}}},\end{equation}
where we include $\gamma_{F\uparrow}\approx 1$ since $c\sim v_{F\downarrow}\gg v_{F\uparrow}$.
Presented assumptions also require $1-\eta\ll1$,
hence we can apply $1+\eta\approx2$.
Numerical solution of equation (\ref{RHD2021ClLM Disp eq u-d}) shows that conditions
$v_{F\downarrow}^{2}k_{z}^{2}/3\gg\omega^{2}\gg v_{F\uparrow}^{2}k_{z}^{2}/3$
cannot be satisfied with the frequency $\omega$ given by equations
(\ref{RHD2021ClLM spectrum seaw small k}) or (\ref{RHD2021ClLM spectrum seaw small k 2}).
Hence, we conclude that equations
(\ref{RHD2021ClLM spectrum seaw small k}) or (\ref{RHD2021ClLM spectrum seaw small k 2})
give a rough analytical illustration of spectrum.
Real frequency of SEAW is considerably higher then given by equations
(\ref{RHD2021ClLM spectrum seaw small k}) or (\ref{RHD2021ClLM spectrum seaw small k 2}).

Approximate analysis of the SEAWs presented above is made as the analytical illustration of relativistic effects in the spectrum of the SEAWs.
Complete analysis is made numerically with the account of the motion of ions.
Therefore, the dispersion equation has the following form:
$$1=\frac{1}{\gamma_{F\uparrow}}\frac{\omega_{L\uparrow}^{2}}{\omega^{2}-u_{p\uparrow}^{2}k_{z}^{2}}$$
\begin{equation}\label{RHD2021ClLM Disp eq u-d-i}
+\frac{1}{\gamma_{F\downarrow}}\frac{\omega_{L\downarrow}^{2}}{\omega^{2}-u_{p\downarrow}^{2}k_{z}^{2}}
+\frac{1}{\gamma_{Fi}}\frac{\omega_{Li}^{2}}{\omega^{2}-u_{pi}^{2}k_{z}^{2}}, \end{equation}
where
the three wave spectrum is presented, it includes the Langmuir wave, SEAW, and ion-acoustic wave.

Numerical analysis of dispersion equation (\ref{RHD2021ClLM Disp eq u-d-i}) is presented in Figs.
(\ref{RHD2021ClLM Fig 01}), (\ref{RHD2021ClLM Fig 02}), (\ref{RHD2021ClLM Fig 03}), and (\ref{RHD2021ClLM Fig 04}).
We consider two regimes of the relatively large spin polarizations $\eta=0.1$ and $\eta=0.9$ to give distinctive illustration of the spin effects.
Fig. (\ref{RHD2021ClLM Fig 01}) shows spectra for three different concentrations at the fixed spin polarization $\eta=0.1$.
We apply the dimensionless parameter $a=(3\pi^{2}n_{0e})^{1/3}\hbar/m_{e}c$
which characterize the role of large concentration and the role of the relativistic effects.
Value of $a$ above 1 corresponds to the noticeable relativistic effects.
Concentrations are chosen to show the spectra in the weakly relativistic regime $a=0.1$ (the upper figure in Fig. (\ref{RHD2021ClLM Fig 01})),
the considerable relativistic effects $a=1$ (the middle figure in Fig. (\ref{RHD2021ClLM Fig 01})),
the strong relativistic effects $a=10$ (the lower figure in Fig. (\ref{RHD2021ClLM Fig 01})).
In the weakly relativistic regime $a=0.1$
we find that the SEAW is a low frequency wave,
but its frequency is well above the frequency of the ion-acoustic wave.
The Langmuir wave in this regime has the well-known non-relativistic form,
but in small $k$ limit its frequency becomes smaller then $\omega_{Le}$.
It is the manifestation of small relativistic effects.
Increase of concentration leads to the increase of the Fermi velocity,
hence the phase velocities of waves increase either.
We see this effect in the middle and lower figures in Fig. (\ref{RHD2021ClLM Fig 01}).
Moreover, the increase of the relativistic effects gives the decrease of minimal frequency of the Langmuir wave.
Same effect is found for the relativistically hot plasmas \cite{Andreev 2021 05}.
Further increase of concentration leads to the ultra relativistic regime demonstrated in Fig. (\ref{RHD2021ClLM Fig 02}).
Simultaneous increase of the relativistic effects and phase velocity of waves leads
to rapprochement of the dispersion dependencies of the Langmuir wave and SEAW.
However, on the smaller scale their frequencies have considerable difference (see the lower figure in Fig. (\ref{RHD2021ClLM Fig 02})).
Larger spin effects $\eta=0.9$ leads to larger differences of frequencies of the Langmuir wave and SEAW
(see Figs. (\ref{RHD2021ClLM Fig 03}) and Fig. (\ref{RHD2021ClLM Fig 04})).
The phase velocity of waves decreases under influence of the relativistic effects at the fixed concentration.
However, we change the concentration during the numerical analysis.
Therefore, the increase of phase velocity at the increase of concentration hide this effect in figures.

\subsection{Small amplitude spin-electron-acoustic soliton}

In this section we consider the nonlinear small amplitude dynamics of the spin-electron-acoustic waves,
which leads to the spin-electron-acoustic solitons in the high-density low-temperature electron-ion plasmas.
Technically this analysis shows similarity to the model of the ion-acoustic solitons described in Ref. \cite{Andreev 2021 11}.
However, the final equations give different physical picture due to the account of the separate spin evolution of electrons.

\begin{figure} \includegraphics[width=8cm,angle=0]{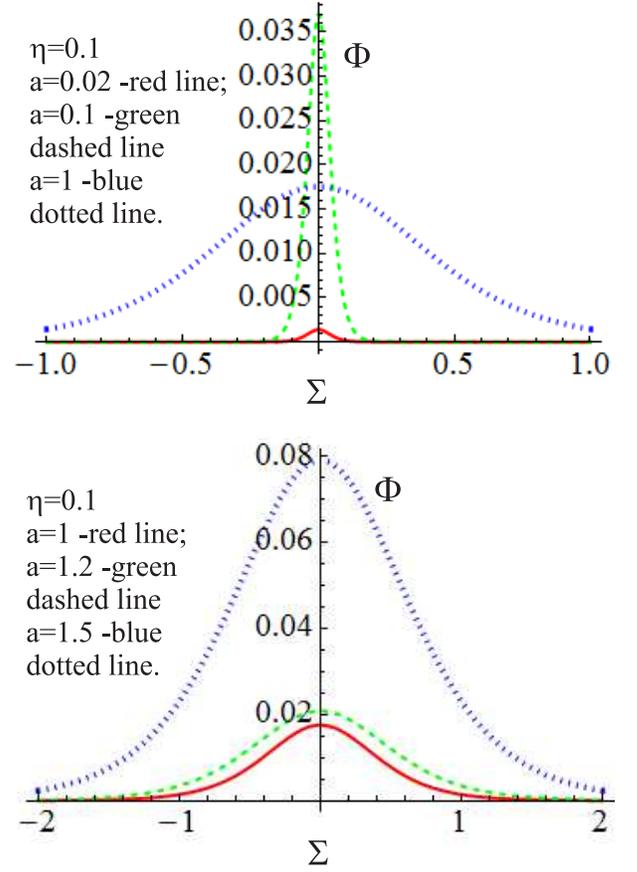}
\caption{\label{RHD2021ClLM Fig 05}
The form of the scalar potential of the electric field in the spin-electron-acoustic soliton
is demonstrated for different concentrations presented via parameter $a=(3\pi^{2}n_{0e})^{1/3}\hbar/m_{e}c$ at fixed spin polarization $\eta=0.1$.
The dimensionless parameters defining the profile of soliton have following definitions
$\Phi=e\varphi_{1}/\tilde{u}p_{Fe}$
and
$\Sigma=(m_{e}\omega_{Le}\sigma/p_{Fe})(\sqrt{m_{e}\tilde{u}/p_{Fe}})$.
}\end{figure}

\begin{figure} \includegraphics[width=8cm,angle=0]{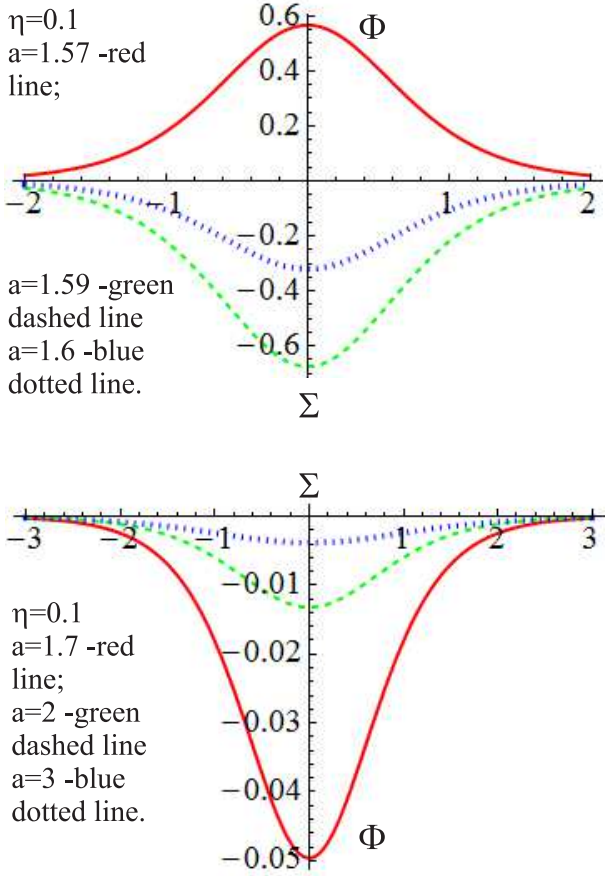}
\caption{\label{RHD2021ClLM Fig 06}
The form of the scalar potential of the electric field in the spin-electron-acoustic soliton
is demonstrated for different concentrations presented via parameter $a=(3\pi^{2}n_{0e})^{1/3}\hbar/m_{e}c$ at fixed spin polarization $\eta=0.1$.
}\end{figure}

In order to study the SEAWs
we introduce the following scaling \cite{Andreev_Iqbal PoP 16}
\begin{equation}\label{RHD2021ClLM def of xi}\begin{array}{cc}
\xi=\varepsilon^{\frac{1}{2}}(z-Ut), & \tau=\varepsilon^{\frac{3}{2}}t,  \end{array} \end{equation}
where
parameter the parameter $\tau$ is the slow time,
while faster dependence on time $t$ is presented via the parameter $\xi$.

Next, we use the following expansions of the hydrodynamic functions on the small parameter $\varepsilon$:
\begin{equation}\label{RHD2021ClLM expansion of n s}  n_{s}=n_{0s}+\varepsilon n_{1s}+\varepsilon^{2} n_{2s},\end{equation}
\begin{equation}\label{RHD2021ClLM expansion of v s}  v_{sz}=0+\varepsilon v_{1sz}+\varepsilon^{2} v_{2sz},\end{equation}
\begin{equation}\label{RHD2021ClLM expansion of Gamma}  \Gamma_{s}=\Gamma_{0s}+\varepsilon \Gamma_{1s}+\varepsilon^{2} \Gamma_{2s},\end{equation}
\begin{equation}\label{RHD2021ClLM expansion of t flux of G}  t_{sz}=0+\varepsilon t_{1sz}+\varepsilon^{2} t_{2sz},\end{equation}
and
\begin{equation}\label{RHD2021ClLM expansion of phi} \phi=0+\varepsilon \phi_{1}+\varepsilon^{3} \phi_{2},\end{equation}
where
subindex $s$ corresponds to $s=i$ for ions, $s=\uparrow$ for the spin-up electrons, $s=\downarrow$ for the spin-down electrons,
function $\Gamma_{0s}$ is given by the equilibrium equation of state (\ref{RHD2021ClLM Gamma rel eq of state}),
and
$\phi$ is the potential of the electric field $\textbf{E}=-\nabla\phi$.

Equations of state (\ref{RHD2021ClLM p rel eq of state})-(\ref{RHD2021ClLM M rel eq of state})
for functions $p_{s}$, $t_{s}$, and $M_{s}$ are applied to get representations of these functions
via $n_{0s}$, $n_{1s}$, $n_{1s}^{2}$, and $n_{2s}$.
For the flux of current of particles we obtain:
\begin{equation}\label{RHD2021ClLM expansion of p on epsilon}
\tilde{p}_{s}\approx \tilde{p}_{0s}+\varepsilon u_{ps}^{2} n_{1s}
+\varepsilon^{2} u_{ps}^{2} n_{2s} +\varepsilon^{2} \frac{v_{Fs}^{2}}{9\gamma_{Fs}^{2}n_{0s}} n_{1s}^{2},
\end{equation}
where $u_{ps}^{2}=v_{Fs}^{2}/3$.
Second, we present the expansion of the partial
fluxes of the reverse relativistic gamma factors
\begin{equation}\label{RHD2021ClLM expansion of p on epsilon}
\tilde{t}_{s}\approx \tilde{t}_{0s}+\varepsilon \frac{v_{Fs}^{2}}{3\gamma_{Fs}} n_{1s}.
\end{equation}
We need it up to the first order on $\varepsilon$.
We do not need to consider the expansion for function $M_{s0}$,
since we need their equilibrium expressions only.

The continuity equations for partial concentrations of electrons and for the concentration of ions
considered in the first (lowest) order of the expansion is
\begin{equation}\label{RHD2021ClLM cont eq expansion order 1} n_{0s}\partial_{\xi}v_{sx1}=U\partial_{\xi}n_{s1}, \end{equation}
while in the second order of expansion we obtain
\begin{equation}\label{RHD2021ClLM cont eq expansion order 2}
\partial_{\tau}n_{s1}-U\partial_{\xi}n_{s2}+\partial_{\xi}(n_{s0}v_{sz2}+n_{s1}v_{sz1})=0. \end{equation}
During expansion of equations we obtain coefficients in front of $\varepsilon^{3/2}$ and $\varepsilon^{5/2}$
for the first and second orders of expansion, correspondingly.

We can integrate equation (\ref{RHD2021ClLM cont eq expansion order 1}).
We use the boundary condition,
where the perturbation caused by soliton goes to zero at infinite distance from its center
$v_{sx1}\rightarrow0$ and $n_{s1}\rightarrow0$ at $\xi\rightarrow\pm\infty$.
Therefore, we obtain
\begin{equation}\label{RHD2021ClLM cont eq expansion order 1 integrated} n_{0s}v_{sz1}=Un_{s1}. \end{equation}

As the second equation, we consider the expansion of the Poisson equation
\begin{equation}\label{RHD2021ClLM Poisson equation I order} n_{\uparrow1}+n_{\downarrow1}-n_{i1}=0, \end{equation}
and
\begin{equation}\label{RHD2021ClLM Poisson equation II order}
-\partial_{\xi}^{2}\varphi_{1}=4\pi (q_{e}(n_{\uparrow2}+n_{\downarrow2})+q_{i}n_{i2}). \end{equation}
Equation found in the first order (\ref{RHD2021ClLM Poisson equation I order})
does not contain the explicit contribution of the potential of the electric field.
The Poisson equation found in the second order (\ref{RHD2021ClLM Poisson equation II order}) contains
the first order expansion of the potential of the electric field $\varphi_{1}$.

\begin{figure} \includegraphics[width=8cm,angle=0]{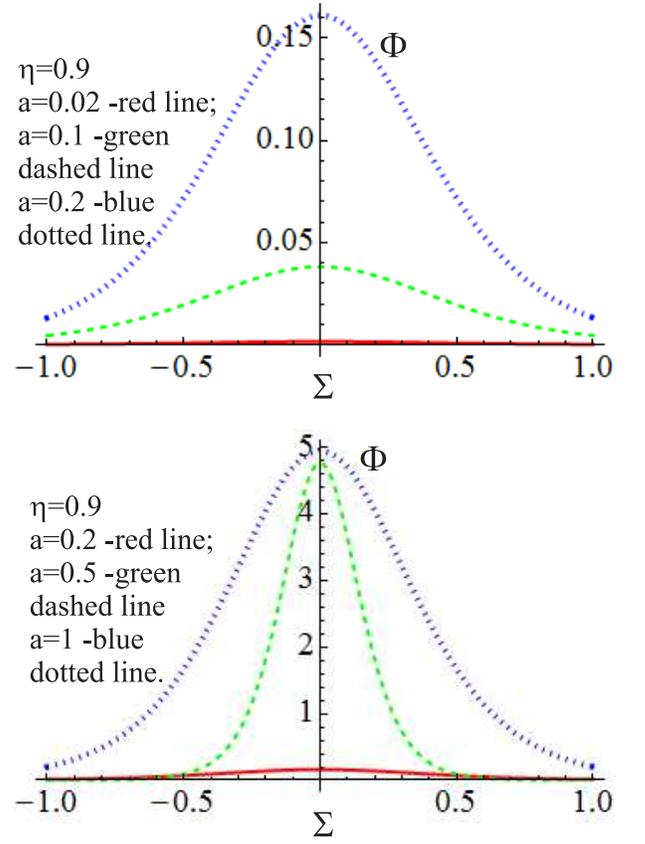}
\caption{\label{RHD2021ClLM Fig 07}
The form of the scalar potential of the electric field in the spin-electron-acoustic soliton
is demonstrated for different concentrations presented via parameter $a=(3\pi^{2}n_{0e})^{1/3}\hbar/m_{e}c$ at fixed spin polarization $\eta=0.9$.
}\end{figure}

\begin{figure} \includegraphics[width=8cm,angle=0]{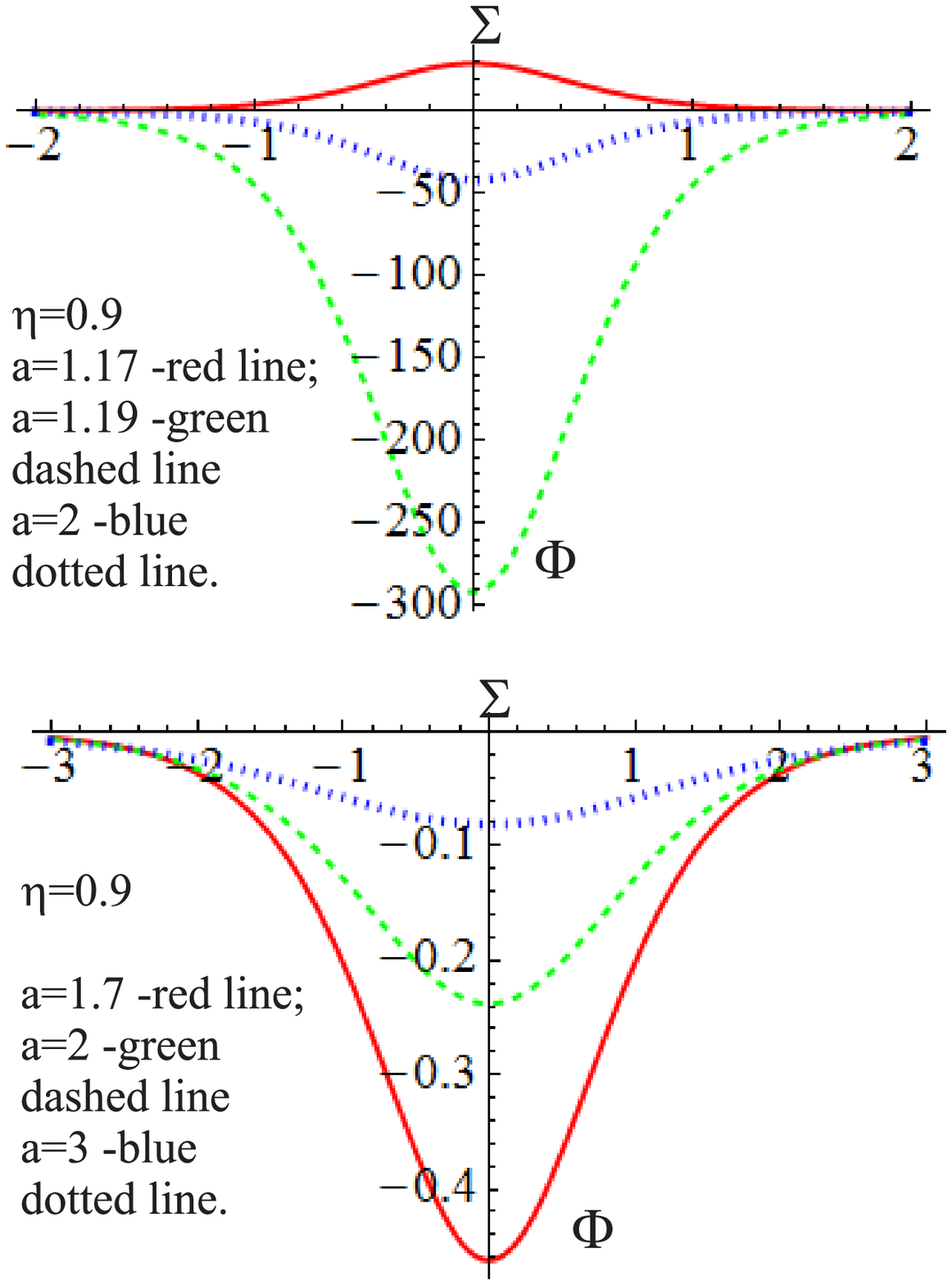}
\caption{\label{RHD2021ClLM Fig 08}
The form of the scalar potential of the electric field in the spin-electron-acoustic soliton
is demonstrated for different concentrations presented via parameter $a=(3\pi^{2}n_{0e})^{1/3}\hbar/m_{e}c$ at fixed spin polarization $\eta=0.9$.
}\end{figure}

The influence of the electric field on the dynamic of electrons and ions can be found from the Euler equation,
which is presented in the first and second orders of expansion:
\begin{equation}\label{RHD2021ClLM Euler equation I order} -Un_{0s}\partial_{\xi}v_{sz1}+u_{sp}^{2}\partial_{\xi}n_{s1}
=-\frac{q_{s}}{m_{s}}n_{0s}\frac{1}{\gamma_{Fs}}\partial_{\xi}\varphi_{1}, \end{equation}
and
$$-Un_{0s}\partial_{\xi}v_{sz2}+n_{0s}\partial_{\tau}v_{sz1}+u_{sp}^{2}\partial_{\xi}n_{s2}+\frac{u_{sp}^{2}}{3\gamma_{Fs}^{2}n_{0s}}\partial_{\xi}n_{s1}^{2}$$
\begin{equation}\label{RHD2021ClLM Euler equation II order} =-\frac{q_{s}}{m_{s}}\biggl(\Gamma_{s1}-\frac{u_{st}^{2}}{c^{2}}n_{s1} \biggr)\partial_{\xi}\varphi_{1}.\end{equation}
The first order Euler equation (\ref{RHD2021ClLM Euler equation I order}) can be integrated.
It gives additional relation between $v_{sz1}$, $n_{s1}$, and $\varphi_{1}$.

The second order Euler equation (\ref{RHD2021ClLM Euler equation II order}) gives relation between $v_{sz2}$, $n_{s2}$, $v_{sz1}$, and $\varphi_{1}$,
but it also includes the first order perturbation of the relativistic hydrodynamic gamma function $\Gamma_{s1}$.
We include equation (\ref{RHD2021ClLM eq for Gamma}) in the lowest order of expansion
\begin{equation}\label{RHD2021ClLM Gamma eq I order}
-U\partial_{\xi}\Gamma_{s1}+\Gamma_{0s}\partial_{\xi}v_{sz1}+\partial_{\xi}t_{sz1}=0. \end{equation}
Equation (\ref{RHD2021ClLM Gamma eq I order}) shows that we need to include equation
for the first order perturbation of the flux of the relativistic hydrodynamic gamma function $t_{sz1}$:
$$U\partial_{\xi}t_{sz1}-u_{st}^{2}\partial_{\xi}n_{s1}+\frac{\Gamma_{0s}}{n_{0s}}u_{sp}^{2}\partial_{\xi}n_{s1}
+\frac{q_{s}}{m_{s}}\Gamma_{0s}
\frac{1}{\gamma_{Fs}}\partial_{\xi}\varphi_{1}$$

\begin{equation}\label{RHD2021ClLM t evol eq I order} =\frac{q_{s}}{m_{s}}n_{0s} \biggl(1-\frac{5u_{sp}^{2}}{c^{2}}+\frac{10}{3}\frac{u_{Ms}^{4}}{c^{4}}\biggr)\partial_{\xi}\varphi_{1}, \end{equation}
where $u_{Ms}^{4}\equiv M_{0s}/n_{0s}$.

In the first order on $\varepsilon$
we obtain the expressions for the perturbation of concentration of each species as the function of the potential of the electric field
\begin{equation}\label{RHD2021ClLM n s1 via phi}
n_{s1}=\frac{q_{s}}{m_{s}\gamma_{Fs}}
\frac{n_{0s}}{U^{2}-u_{sp}^{2}}\varphi_{1}. \end{equation}
We substitute expression (\ref{RHD2021ClLM n s1 via phi}) obtained for each species in the Poisson equation (\ref{RHD2021ClLM Poisson equation I order}).
It gives us equation for the velocity of perturbation $U$:
\begin{equation}\label{RHD2021ClLM eq for U}
\frac{1}{\gamma_{F\uparrow}}
\frac{1}{U^{2}-u_{p\uparrow}^{2}}
+\frac{1}{\gamma_{F\downarrow}}
\frac{1}{U^{2}-u_{p\downarrow}^{2}}
+\frac{m_{e}}{m_{i}\gamma_{Fi}}
\frac{1}{U^{2}-u_{ip}^{2}}=0.\end{equation}
Equation (\ref{RHD2021ClLM eq for U}) can have solutions under condition that
denumerators of different terms have different sings,
for instance $u_{ip}^{2}< U^{2}< u_{ep}^{2}$.
The right-hand side of equation (\ref{RHD2021ClLM eq for U}) is equal to zero.
This condition exclude the Langmuir wave solution.
This equation gives two solutions.
One corresponds to the SEAW and the second solution corresponds to the ion-acoustic wave.

Our analysis of the second order approximation leads to the nonlinear equation for the electric potential
\begin{widetext}
$$\partial_{\xi}^{3}\varphi_{1}
+\sum_{s=\uparrow,\downarrow,i}\frac{U\omega_{Ls}^{2}}{\gamma_{Fs}(U^{2}-u_{sp}^{2})^{2}}
\partial_{\tau}\varphi_{1}
+\sum_{s=\uparrow,\downarrow,i}\frac{q_{s}}{m_{s}}
\frac{2(U^{2}+\frac{u_{sp}^{2}}{3\gamma_{Fs}^{2}})\omega_{Ls}^{2}}{\gamma_{Fs}^{2}(U^{2}-u_{sp}^{2})^{3}}\varphi_{1}\partial_{\xi}\varphi_{1}$$
$$+\sum_{s=\uparrow,\downarrow,i}\frac{q_{s}}{m_{s}}\frac{\omega_{Ls}^{2}}{\gamma_{Fs}(U^{2}-u_{sp}^{2})^{2}}
\biggl[\frac{\Gamma_{0s}}{n_{0s}}\biggl(1-\frac{u_{sp}^{2}}{U^{2}}\biggr)
+\frac{1}{U^{2}}\biggl(\frac{v_{Fs}^{2}}{3\gamma_{Fs}}-u_{st}^{2}\frac{U^{2}}{c^{2}}\biggr)\biggr]\varphi_{1}\partial_{\xi}\varphi_{1}$$
\begin{equation}\label{RHD2021ClLM KdV simm}
+\sum_{s=\uparrow,\downarrow,i}\frac{q_{s}}{m_{s}}\frac{\omega_{Ls}^{2}}{U^{2}(U^{2}-u_{sp}^{2})}
\biggl(1-5\frac{p_{0s}}{n_{0s}c^{2}}-\frac{\Gamma_{0s}}{\gamma_{Fs}n_{0s}}+\frac{10}{3}\frac{u_{sM}^{4}}{c^{4}}\biggr)
\varphi_{1}\partial_{\xi}\varphi_{1}=0. \end{equation}

\end{widetext}

Here we have Korteweg-de Vries (KdV) equation in terms of electrostatic potential.
In schematic form it can be rewritten as
$\partial_{\xi}^{3}\varphi_{1}+D\partial_{\tau}\varphi_{1} +(P/2)\partial_{\xi}\varphi_{1}^{2}=0$,
where we partially follow notations of Ref. \cite{Andreev_Iqbal PoP 16}.
Hence, parameter $D$ corresponds to the coefficient in the second term of equation (\ref{RHD2021ClLM KdV simm}),
while $P$ corresponds to the superposition of coefficients of the third, fourth and fifth terms.
To get the soliton solution of the Korteweg-de Vries equation (\ref{RHD2021ClLM KdV simm}),
we present novel variable
$\sigma=\xi-\tilde{u}\tau$,
where $U$ is the soliton propagation velocity.
Consequently, we find
\begin{equation}\label{RHD2021ClLM solution of KdV}
\varphi_{1}=\frac{3D\tilde{u}}{P} \frac{1}{\cosh^{2}(\sqrt{D\tilde{u}}\sigma/2)}.
\end{equation}

Let us illustrate main features of the spin-electron-acoustic solitons at the numerical analysis presented on Figs.
(\ref{RHD2021ClLM Fig 05}), (\ref{RHD2021ClLM Fig 06}), (\ref{RHD2021ClLM Fig 07}), (\ref{RHD2021ClLM Fig 08}).
Presentation of results is made for two spin polarization regimes similarly to analysis of spectra of small amplitude waves.
Fig. (\ref{RHD2021ClLM Fig 05}) shows the existence of the "bright" soliton,
the soliton of increased potential of the electric field.
The upper figure in Fig. (\ref{RHD2021ClLM Fig 05}) shows that
the increase of concentration from $a=0.02$ up to $a=0.1$
(both regimes corresponds to the small relativistic corrections)
we have increase of amplitude and width of the spin-electron-acoustic soliton.
Further increase of concentration up to $a=1$ (the relativistic regime)
leads to decrease of the amplitude and increase of the width.

The lower figure in Fig. (\ref{RHD2021ClLM Fig 05}) demonstrates further growth of the concentration,
where the increase of the amplitude with the simultaneous increase of width is found.

The upper figure in Fig. (\ref{RHD2021ClLM Fig 06}) presents the following modification of the soliton profile
during the increase of concentration at the fixed spin polarization $\eta=0.1$.
Small increase of concentration from $a=1.5$ to $a=1.57$ demonstrates considerable increase of amplitude from $\Phi=0.08$ to $\Phi=0.55$.
Further small increase of concentration to $a=1.59$ leads to dramatic change of the soliton profile,
so we have dark soliton (area of negative potential of the electric field) instead of the bright soliton.
While module of amplitude shows no considerable changes (there is small increase).
Following small increase of concentration to $a=1.6$ shows considerable decrease of the module of amplitude and width of soliton.
Continuing the increase of concentration up to values $a=1.7$, $a=2$, $a=3$ we find monotonic decrease of amplitude and small decrease of width
(see the lower figure in Fig. (\ref{RHD2021ClLM Fig 06})).
Figs. (\ref{RHD2021ClLM Fig 07}) and (\ref{RHD2021ClLM Fig 08}) show same tendency for another spin polarization $\eta=0.9$,
but the critical point,
where the sign of the amplitude of soliton is changed.
The change happens at $a=1.18$.
So, the critical concentration becomes smaller at larger spin polarization.

\section{Conclusion}

In order to consider spin-electron-acoustic waves and spin-electron-acoustic solitons propagating parallel to the external magnetic field
in the relativistic degenerate partially spin polarized electron gas,
we have presented a generalization of the relativistic
hydrodynamic model with the average reverse gamma factor evolution.
Suggested model has been based on the following background.
The relativistic hydrodynamics with is originally developed for the hot plasmas
with temperature of electrons $T_{e}$ of order of the rest energy $T_{e}\sim m_{e}c^{2}$.
However, this model has a restriction from the area of large concentrations:
$T_{e}\gg T_{Fe}=(3\pi^{2}n_{0e})^{2/3}\hbar^{2}/2m_{e}$.
Modification of hydrodynamic model with the average reverse gamma factor evolution
for the high-density degenerate relativistic electrons $ T_{Fe}\sim m_{e}c^{2}\gg T_{e}$ is developed in literature as well
in the regime of electrons with the zero spin polarization.
The area of applications of the single fluid model of spin polarized electrons has been discussed.

Degenerate electrons is a quantum system.
Proper justification of this model should be based on quantum dynamics of particles.
However, major plasma effects are based on the electromagnetic interaction and form of distribution of particles in the momentum space,
which can be captured in the quasi-classic limit,
as it has been done in this paper.
Other quantum effects like the quantum Bohm potential or dynamic of the transverse projections of the spin density have been neglected.

The dispersion equation for the spin-electron-acoustic wave is considered in two regimes:
\newline
1) immobile ions;
\newline
2) separate spin evolution of electrons with the account of the motion of ions.
\newline
The first regime shows that
the large spin polarization regime leads to the relatively small phase velocity of the spin-electron-acoustic waves,
so the influence of ions is essential.
Similar conclusion is correct in the nonrelativistic regime.
However, the relativistic effects considerably decreases the phase velocity of the spin-electron-acoustic waves.
Numerical analysis of spectra of the Langmuir, spin-electron-acoustic and ion-acoustic waves have been presented.
We stress our attention on the dependence of spectra on the spin polarization of electrons.
The spin polarization of ions is neglected.
The properties of spin-electron-acoustic soliton has been studied.

\section{Acknowledgements}

Work is supported by the Russian Foundation for Basic Research (grant no. 20-02-00476).

\section{DATA AVAILABILITY}

Data sharing is not applicable to this article as no new data were
created or analyzed in this study, which is a purely theoretical one.


\begin{thebibliography}{17}


\bibitem{Andreev PRE 15} P. A. Andreev, "Separated spin-up and spin-down quantum hydrodynamics of degenerated electrons",
Phys. Rev. E \textbf{91}, 033111 (2015).

\bibitem{Andreev AoP 15 SEAW} P. A. Andreev, L. S. Kuz'menkov,
"Oblique propagation of longitudinal waves in
magnetized spin-1/2 plasmas: Independent
evolution of spin-up and spin-down electrons",
Annals of Physics \textbf{361}, 278 (2015).


\bibitem{Andreev PRE 16} P. A. Andreev, Z. Iqbal,
"Rich eight-branch spectrum of the oblique propagating longitudinal waves in partially
spin-polarized electron-positron-ion plasmas",
Phys. Rev. E \textbf{93}, 033209 (2016).



\bibitem{Andreev APL 16} P. A. Andreev, L. S. Kuz'menkov,
"Surface spin-electron acoustic waves in magnetically ordered metals",
Appl. Phys. Lett. \textbf{108}, 191605 (2016).




\bibitem{Andreev EPL 16} P. A. Andreev, L. S. Kuz'menkov,
"Separated spin-up and spin-down evolution of degenerated electrons in two-dimensional
systems: Dispersion of longitudinal collective excitations in plane and nanotube geometry",
Eur. Phys. Lett. \textbf{113}, 17001 (2016).



\bibitem{Andreev_Iqbal PoP 16} Z. Iqbal, P. A. Andreev,
"Nonlinear separate spin evolution in degenerate electron-positron-ion plasmas",
Phys. Plasmas \textbf{23}, 062320 (2016).



\bibitem{Iqbal Shahid PP 17} M. Shahid, Z. Iqbal, M. Jamil, and G. Murtaza,
"Raman three-wave interaction for the O-mode, Shear Alfven wave and the electron plasma perturbations",
Phys. Plasmas \textbf{24}, 102113 (2017).

\bibitem{Iqbal PLA 18} Z. Iqbal, and G. Murtaza,
"Electrostatic solitary structure of the SEAWs is studied at the oblique propagation of the weakly-nonlinear waves",
Phys. Lett A \textbf{382}, 44 (2018).



\bibitem{Trukhanova PLA 15} M. I. Trukhanova,
"",
Phys. Lett. A \textbf{379}, 2777 (2015).





\bibitem{Iqbal PP 18 UHW} Z. Iqbal, I. A. Khan, and G. Murtaza,
"On the upper hybrid wave instability in a spin polarized degenerate plasma",
Phys. Plasmas \textbf{25}, 062121 (2018).
%


\bibitem{Iqbal PP 18 LI} Z. Iqbal, M. Jamil, and G. Murtaza,
"Langmuir instability in partially spin polarized bounded degenerate plasma",
Phys. Plasmas \textbf{25}, 042106 (2018).
%

\bibitem{Andreev PP 16 SSE kin} P. A. Andreev,
"Spin-electron acoustic waves: The Landau damping and ion contribution in the
spectrum",
Phys. Plasmas \textbf{23}, 062103 (2016).

\bibitem{Andreev PoP non-triv kinetics 17} P. A. Andreev, L. S. Kuz'menkov,
"Dielectric permeability tensor and linear waves in spin-1/2 quantum kinetics with nontrivial
equilibrium spin-distribution functions",
Phys. Plasmas \textbf{24}, 112108 (2017).


\bibitem{Andreev PoP kinetics 17 a} P. A. Andreev,
"Kinetic analysis of spin current contribution to spectrum of electromagnetic waves in
spin-1/2 plasma. I. Dielectric permeability tensor for magnetized plasmas",
Phys. Plasmas \textbf{24}, 022114 (2017).

\bibitem{Andreev PoP kinetics 17 b} P. A. Andreev,
"Kinetic analysis of spin current contribution to spectrum of electromagnetic waves in
spin-1/2 plasma. II. Dispersion dependencies",
Phys. Plasmas \textbf{24}, 022115 (2017).


\bibitem{Maksimov QHM 99} L. S. Kuz'menkov, S. G. Maksimov,
"Quantum hydrodynamics of particle systems with
Coulomb interaction and quantum Bohm potential,"
Theor. Math. Phys. \textbf{118}, 227 (1999).

\bibitem{MaksimovTMP 2001} L. S. Kuz'menkov, S. G. Maksimov, V. V. Fedoseev,
"Microscopic quantum hydrodynamics of systems of
fermions: part I",
Theor. Math. Phys. \textbf{126}, 110 (2001).





\bibitem{Marklund PRL07} M. Marklund, G. Brodin,
"Dynamics of Spin-1/2 Quantum Plasmas",
Phys. Rev. Lett. \textbf{98}, 025001 (2007).



\bibitem{Brodin NJP 07} G. Brodin, M. Marklund,
"Spin magnetohydrodynamics",
New J. Phys. \textbf{9}, 277 (2007).




\bibitem{Brodin PRL 08 Cl Reg} G. Brodin, M. Marklund, and G. Manfredi,
"Quantum Plasma Effects in the Classical Regime",
Phys. Rev. Lett. \textbf{100}, 175001 (2008).

%
\bibitem{Brodin PRL 08 g Kin} G. Brodin, M. Marklund, J. Zamanian, A. Ericsson, and P. L. Mana,
"Effects of the g factor in semiclassical kinetic plasma theory",
Phys. Rev. Lett. \textbf{101}, 245002 (2008).

%
\bibitem{Brodin PRL 10} G. Brodin, A. P. Misra, and M. Marklund,
"Spin Contribution to the Ponderomotive Force in a Plasma",
Phys. Rev. Lett. \textbf{105}, 105004 (2010).

\bibitem{Shukla PhUsp 2010} P. K. Shukla, B. Eliasson,
"Nonlinear aspects of quantum plasma physics ",
Phys. Usp. \textbf{53}, 51 (2010).

\bibitem{Shukla RMP 11} P. K. Shukla, B. Eliasson,
"Nonlinear collective interactions in quantum plasmas with degenerate electron fluids",
Rev. Mod. Phys. \textbf{83}, 885 (2011).

\bibitem{Mahajan PRL 11} S. M. Mahajan, F. A. Asenjo,
"Vortical Dynamics of Spinning Quantum Plasmas: Helicity Conservation",
Phys. Rev. Lett. \textbf{107}, 195003 (2011).



\bibitem{Koide PRC 13} T. Koide,
"Spin-electromagnetic hydrodynamics and magnetization induced by spin-magnetic interaction",
Phys. Rev. C \textbf{87}, 034902 (2013).

\bibitem{Andreev PP 15 Positrons} P. A. Andreev,
"Hydrodynamic and kinetic models for spin-1/2 electron-positron quantum plasmas:
Annihilation interaction, helicity conservation, and wave dispersion in magnetized
plasmas",
Phys. Plasmas \textbf{22}, 062113 (2015).



\bibitem{Yoshida JPA 16} Z. Yoshida, S. M. Mahajan,
"Quantum spirals",
J. Phys. A: Math. Theor. \textbf{49}, 055501 (2016).


\bibitem{Uzdensky RPP review 14} D. A. Uzdensky, S. Rightley,
"Plasma physics of extreme astrophysical environments",
Rep. Progr. Phys. \textbf{77}, 036902 (2014).



\bibitem{Dodin PRA 15 First-principle} D. E. Ruiz, I. Y. Dodin,
"First-principle variational formulation of polarization effects in geometrical optics",
Phys. Rev. A \textbf{92}, 043805 (2015).
%
%
\bibitem{Ekman PRE 17} R. Ekman, F. A. Asenjo, and J. Zamanian,
"Relativistic kinetic equation for spin-1/2 particles in the long-scale-length approximation",
Phys. Rev. E \textbf{96}, 023207 (2017).


\bibitem{Andreev Ch 21} P. A. Andreev,
"Quantum hydrodynamic theory of quantum fluctuations in dipolar Bose–-Einstein condensate",
Chaos \textbf{31}, 023120 (2021).

\bibitem{Andreev PoF 21} P. A. Andreev, I. N. Mosaki, and M. I. Trukhanova,
"Quantum hydrodynamics of the spinor Bose–Einstein condensate at non-zero temperatures",
Phys. Fluids \textbf{33}, 067108 (2021).
10.1063/5.0053035


\bibitem{Andreev JPP 21} P. A. Andreev,
"Hydrodynamics of quantum corrections to the Coulomb interaction via the third rank tensor
evolution equation: application to Langmuir waves and spin-electron acoustic waves",
J. Plasma Phys. \textbf{87}, 905870511 (2021).

\bibitem{Andreev 2021 11} P. A. Andreev,
"Relativistic hydrodynamic model with the average reverse gamma factor evolution for the degenerate plasmas:
high-density ion-acoustic solitons",
arXiv:2111.14197.

\bibitem{Andreev 2021 05} P. A. Andreev,
"On the structure of relativistic hydrodynamics for hot plasmas", arXiv:2105.10999.

\bibitem{Andreev 2021 09} P. A. Andreev,
"Microscopic model for relativistic hydrodynamics of ideal plasmas",
arXiv:2109.14050.

\bibitem{Andreev 2021 10} P. A. Andreev,
"Anisotropic pressure effects in hydrodynamic description of waves propagating parallel to the magnetic field in relativistically hot plasmas",
arXiv:2110.14749.



\bibitem{Andreev 2021 06} P. A. Andreev,
"Waves propagating parallel to the magnetic field in relativistically hot plasmas: A hydrodynamic models", arXiv:2106.14327.

\bibitem{Andreev 2021 07} P. A. Andreev,
"On a hydrodynamic description of waves propagating perpendicular to the magnetic field in relativistically hot plasmas",
arXiv:2107.13603.

\bibitem{Andreev 2021 08} P. A. Andreev,
"A hydrodynamic model of Alfvenic waves and fast magneto-sound in the
relativistically hot plasmas at propagation parallel to the magnetic field",
arXiv:2108.12721.






\bibitem{Kuz'menkov 91} L. S. Kuz'menkov,
"Field form of dynamics and statistics of systems of particles with electromagnetic interaction",
Theoretical and Mathematical Physics \textbf{86}, 159 (1991).


\bibitem{Drofa TMP 96}  M. A. Drofa, L. S. Kuz'menkov,
"Continual approach to multiparticle systems with long-range interaction. Hierarchy of macroscopic fields and physical consequences",
Theoretical and Mathematical Physics \textbf{108}, 849 (1996).



\bibitem{Andreev PIERS 2012} L. S. Kuz'menkov and P. A. Andreev,
"Microscopic Classic Hydrodynamic and Methods of Averaging",
presented in PIERS Proceedings, p. 158, Augoust 19-23, Moscow,
Russia 2012.



\bibitem{Asenjo PP 11} F. A. Asenjo, V. Munoz, J. A. Valdivia, and S. M. Mahajan,
"A hydrodynamical model for relativistic spin quantum plasmas",
Phys. Plasmas \textbf{18}, 012107 (2011).



\bibitem{Ryan PRB 91} J. C. Ryan,
"Collective excitations in a spin-polarized quasi-two-dimensional electron gas",
Phys. Rev. B \textbf{43}, 4499 (1991).



\bibitem{Agarwal PRL 11} A. Agarwal, M. Polini, R. Fazio, G. Vignale,
"Persistent Spin Oscillations in a Spin-Orbit-Coupled Superconductor",
Phys. Rev. Lett. \textbf{107}, 077004 (2011).

\bibitem{Agarwal PRB 14} A. Agarwal, M. Polini, G. Vignale, M. E. Flatte,
"Long-lived spin plasmons in a spin-polarized two-dimensional electron gas",
Phys. Rev. B \textbf{90}, 155409 (2014).


\bibitem{Perez PRB 09} F. Perez,
"Spin-polarized two-dimensional electron gas embedded in a semimagnetic quantum well: Ground state, spin responses, spin excitations, and Raman spectrum",
Phys. Rev. B \textbf{79}, 045306 (2009).










\bibitem{Hakim book Rel Stat Phys} Remi Hakim,
Introduction to Relativistic Statistical Mechanics Classical and Quantum, World Scientific Publishing Co. Pte. Ltd.,  2011.




\bibitem{Shatashvili ASS 97} N. L. Shatashvili, J. I. Javakhishvili, H. Kaya,
"Nonlinear wave dynamics in two-temperature electron-positron-ion plasma",
Astrophys Space Sci. \textbf{250}, 109 (1997).

\bibitem{Mahajan PRL 03} S. M. Mahajan,
"Temperature-Transformed "Minimal Coupling": Magnetofluid Unification",
Phys. Rev. Lett. \textbf{90}, 035001 (2003).


\bibitem{Mahajan PoP 2011} S. M. Mahajan, Z. Yoshida,
"Relativistic generation of vortex and magnetic field",
Phys. Plasmas \textbf{18}, 055701 (2011).


\bibitem{Heyvaerts AA 12}
J. Heyvaerts, T. Lehner, and F. Mottez,
"Non-linear simple relativistic Alfven waves in astrophysical plasmas",
A\&A \textbf{542}, A128 (2012).


\bibitem{Asenjo 19} F. A. Asenjo, and L. Comisso,
"Gravitational electromotive force in magnetic reconnection around Schwarzschild black holes",
Phys. Rev. D \textbf{99}, 063017 (2019).



\bibitem{Shatashvili PoP 20} N. L. Shatashvili, S. M. Mahajan , and V. I. Berezhiani,
"Nonlinear coupling of electromagnetic and electron acoustic waves in multi-species degenerate astrophysical plasma",
Phys. Plasmas \textbf{27}, 012903 (2020).

\bibitem{Comisso PRL 14} L. Comisso, F. A. Asenjo,
"Thermal-Inertial Effects on Magnetic Reconnection in Relativistic Pair Plasmas",
Phys. Rev. Lett. \textbf{113}, 045001 (2014).


\bibitem{Liu PPCF 21} Z. Y. Liu, Y. Z. Zhang, and S. M. Mahajan,
"The effect of curvature induced broken potential
vorticity conservation on drift wave turbulences",
Plasma Phys. Control. Fusion \textbf{63}, 045009 (2021).



\bibitem{Comisso 19} L. Comisso, and F. A. Asenjo,
"Generalized Magnetofluid Connections in Curved Spacetime",
Phys. Rev. D \textbf{102}, 023032 (2020).


\bibitem{Mahajan PoP 16} S. M. Mahajan, F. A. Asenjo,
"A statistical model for relativistic quantum fluids interacting with an intense electromagnetic wave",
Phys. Plasmas \textbf{23}, 056301 (2016).

\bibitem{Brunetti MNRAS 04}
G. Brunetti, P. Blasi, R. Cassano, and S. Gabici,
"Alfvenic reacceleration of relativistic particles in galaxy clusters: MHD waves, leptons and hadrons",
Mon. Not. R. Astron. Soc. \textbf{350}, 1174–1194 (2004). 


\bibitem{Mahajan IJTP 14} S. M. Mahajan, F. A. Asenjo,
"Hot Fluids and Nonlinear Quantum Mechanics",
Int. J. Theor. Phys. \textbf{54}, 1435 (2014).


\bibitem{Mendonca PP 11} J. T. Mendonca,
"Wave kinetics of relativistic quantum plasmas",
Phys. Plasmas \textbf{18}, 062101 (2011).

\bibitem{Zhu PPCF 12} J. Zhu and P. Ji,
"Dispersion relation and Landau damping of waves in high-energy density plasmas",
Plasma Phys. Control. Fusion \textbf{54}, 065004 (2012).
%


\bibitem{Munoz EPS 06}
V. Munoz, T. Hada, and S. Matsukiyo,
"Kinetic effects on the parametric decays of Alfven waves in relativistic pair plasmas",
Earth Planets Space, \textbf{58}, 1213–1217, (2006).



\end{thebibliography}
\end{document}